\documentclass[english]{svjour3}
\usepackage[T1]{fontenc}
\usepackage{xcolor}
\usepackage{graphicx}
\usepackage[authoryear]{natbib}
\PassOptionsToPackage{normalem}{ulem}
\usepackage{ulem}

\makeatletter

\providecolor{lyxadded}{rgb}{0,0,1}
\providecolor{lyxdeleted}{rgb}{1,0,0}

\smartqed  

\makeatother

\usepackage{babel}

\begin{document}

\title{A High Speed Networked Signal Processing Platform for Multi-element
Radio Telescopes}

\author{Peeyush Prasad \and C.R. Subrahmanya}

\institute{Peeyush Prasad \at Astronomy and Astrophysics Group,\\ Raman Research
Institute,\\ Bangalore - 560 080, India\\ Tel: +91-80-23610122\\
Fax: +91-80-23610492 \\ \email{peeyush@rri.res.in} \and C.R. Subrahmanya\at
Astronomy and Astrophysics Group,\\ Raman Research Institute,\\
Bangalore - 560 080, India\\Tel: +91-80-23610122\\Fax: +91-80-23610492
\\\email{crs@rri.res.in}\date{Received: /Accepted:}}
\maketitle
\begin{abstract}
A new architecture is presented for a Networked Signal Processing
System (NSPS) suitable for handling the real-time signal processing
of multi-element radio telescopes. In this system, a multi-element
radio telescope is viewed as an application of a \emph{multi-sensor,
data fusion} problem which can be decomposed into a general set of
computing and network components for which a practical and scalable
architecture is enabled by current technology. The need for such a
system arose in the context of an ongoing program for reconfiguring
the Ooty Radio Telescope (ORT) as a programmable 264-element array,
which will enable several new observing capabilities for large scale
surveys on this mature telescope. For this application, it is necessary
to manage, route and combine large volumes of data whose real-time
collation requires large I/O bandwidths to be sustained. Since these
are general requirements of many multi-sensor fusion applications,
we first describe the basic architecture of the NSPS in terms of a
\emph{Fusion Tree} before elaborating on its application for the ORT.
The paper addresses issues relating to high speed distributed data
acquisition, Field Programmable Gate Array (FPGA) based peer-to-peer
networks supporting significant on-the fly processing while routing,
and providing a last mile interface to a typical commodity network
like Gigabit Ethernet. The system is fundamentally a pair of two co-operative
networks, among which one is part of a commodity high performance
computer cluster and the other is based on Commercial-Off The-Shelf
(COTS) technology with support from software/firmware components in
the public domain.\keywords{Radio Interferometry \and Multibeam Phased Array \and FPGA \and Signal Processing \and High speed networks \and Distributed real-time processing.}
\end{abstract}

\section{Introduction}

A Multi-element Radio Telescope is a spatially spread array of antennas
(or antenna elements) whose noise-like responses are required to be
time aligned, dynamically calibrated and combined or correlated in
real time. The resulting estimates of spatio-temporal and spectral
correlations between responses of pairs of elements can be used to
recover the desired information on the strength and distribution of
radio emission within the common field of view \citep{key-6} using
standard post-processing software. Thus the signal of interest is
statistical in nature, resulting from a minute level of mutual coherence
arising from weak celestial signals buried in noise. Because of the
large number of elements and the high sampling rates necessary for
bandwidths exceeding several tens of MHz in recent arrays, real-time
statistical estimation is essential to achieve practical data rates
and volumes for recording and post processing. For instance, the recently
initiated upgrade for the Ooty Radio Telescope (ORT) aims at treating
the 30m x 506m parabolic cylindrical antenna of the ORT as 264 independent
sets of elements, each of which is required to be sampled at 80 Ms/s,
leading to a data generation rate of 21 Gigasamples per second, exceeding
80 Terabytes per hour. Till recently, computing requirements of this
scale forced a choice of custom hardware to be the most favored platform.
However, rapid developments in the fields of digital technology, communication
and computing have led to a changing trend towards alternative approaches
for upcoming telescopes. Such approaches range between a customized
and reusable hardware library of components on an FPGA platform, e.g.,
the CASPER project \citep{key-2}, and a software-only approach, e.g.,
at the GMRT \citep{key-3}. The GMRT case is an example of a recent
transition from custom hardware to a software-only approach.

In this paper, we have taken a middle path, where the real-time processing
of a multi-element radio telescope is abstracted as a multi-sensor,
data fusion problem and addressed in a new platform called the Networked
Signal Processing System (NSPS) in terms of packetized, heterogeneous,
distributed and real-time signal processing. It is a co-operative
of two kinds of networks, among which one is a custom peer-to-peer
network while the other is a part of a commodity processor network.
The custom network includes subsystems related to the digitization
and all intermediate routing and preprocessing blocks as the network
nodes, in which the emphasis is on traffic shaping, on-the-fly processing
and load balancing for effective distributed computing. However, all
customized protocols are absorbed while crossing over the last mile
to interface to the commodity processor network using a common industry-standard
network protocol. The actual estimation of the correlations is carried
out by nodes on the commodity network.

In contrast to the traditional use of a packetized network as merely
a data transport fabric between processing entities, we have the notion
of {}``a logical packet'' based on an application specific {}``transaction
unit'', which itself may be composed of a large number of physical
packets whose sizes are network-specific This unit refers to a time
stretch long enough to facilitate a dynamical flagging mechanism or/and
to relax the constraint on timing, synchronization and scheduling
of workloads on the commodity Operating Systems on which processing
is expected to be carried out. Both these requirements necessitate
lower level NSPS nodes to be equipped with large memories, which are
also used to route traffic selectively (traffic shaping) to higher
levels in the NSPS. Two independent considerations have led us towards
stretching the transaction unit to a good fraction of a second. One
of these, as explained above, is to provide latency tolerance in order
to simplify software on standard computing platforms, while the other
arises from a desire to make explicit provision for preprocessing
using concepts related to modern Information Theory. From this point
of view and to attract the attention of experts from outside the field
of radio astronomy, we have given a somewhat unconventional description
of the signal path and analysis of processing requirements in Sections
\ref{sec:MultiElementRadioTelescope} and \ref{sec:RealTimeProcessing}
in an attempt to illustrate the connection of the problem to Information
Theory. 

Another concept in literature which we find useful in the present
context is that of {}``multi-sensor data fusion'', defined by \citep{key-1-1},
as a system model where {}``spatially and temporally indexed data
provided by different sources are combined (fused) in order to improve
the processing and interpretation of these data''. This model, widely
used in applications like military target tracking, weather forecasting
etc, has many features relevant for describing the control flow and
pre-processing required in a multi-element radio telescope before
correlation. In a sense, the NSPS is an adaptation of data fusion
architecture to our domain. Our analysis of the nature of the real-time
problem results in a natural partitioning into two broad categories
as elaborated in Section \ref{sec:RealTimeProcessing}. This is our
primary motivation for defining the architecture, described as a Fusion
Tree in Section \ref{sec:FusionTree}. An illustration of the feasibility
of its implementation is presented in Section \ref{sec:An-Implementation-Example:}
by considering the case of the ORT upgrade.

\section{\emph{\label{sec:MultiElementRadioTelescope}}A Multi-Element Radio
Telescope as a Set of Correlated, Noisy Carriers of Information}

We present here an abstract picture of a multi-element radio telescope,
in which the primary beams of individual elements are viewed as virtual
communication channels carrying different combinations of radiation
from a set of independent celestial {}``radio emitters''. These
are located in different directions within the common celestial region
intercepted by the element primary beams. Each individual radio emitter
from this set is a source of stationary (within the observing time)
random process. Since signals received in any finite bandwidth can
be spectrally decomposed, each primary beam can be considered equivalent
to a set of independent communication channels of identical bandwidths
corresponding to the spectral resolution. Each such channel is characterized
by a correlation timescale equal to reciprocal of its bandwidth. Thus,
spectral decomposition can be used to enable the propagation time
differences for noise from different radio emitters to be within their
correlation timescale. This implies that the corresponding channels
of different primary beams are carriers of noise arising from different
superpositions of the same set of random processes, but with different
weights. Information about the strength and distribution of radio
emitters in the field of view can be considered to be coded in the
correlation between corresponding communication channels of different
elements. Hence, we consider the real-time spectral correlation to
be a fusion process for compressing the information conveyed by the
responses of different elements without affecting decoding to be carried
out in post-processing, say in the form of an image of a celestial
region. 

\begin{figure}
\center\includegraphics[bb=0bp 0bp 990bp 794bp,scale=0.55]{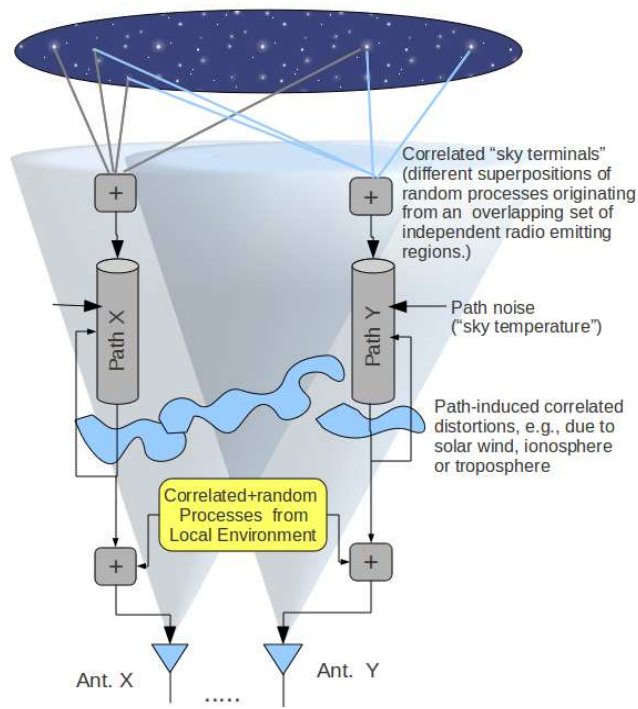}\caption{\emph{\label{fig:Antenna-beams-viewed}Antenna beams viewed as noisy
virtual communication channels connecting a sky region and an antenna
element.}}

\end{figure}

However, in practice, the propagation medium introduces a variety
of correlated and uncorrelated noise into these virtual channels which
can erase or distort the information related to celestial emitters.
This is represented schematically in Fig. \ref{fig:Antenna-beams-viewed}.
Often, some of these distortions are characterized by a combination
of dispersive and non-dispersive components which are\textit{ localized}
in time and/or frequency, unlike the celestial signal which is more
like a random noise. Such a localization can be exploited by a pre-processing
algorithm to enable suitable recognition and characterization of deviant
(non-random) data and to tag them suitably. Such data can then be
segregated from those passed on to an irreversible fusion operation,
e.g., to minimize the biases in the correlations. On the other hand,
in situations like observing fast transients with low duty cycle,
or when one wants to find the direction in which such a non-random
signal is present, the segregated data can be passed to an independent
processing or recording stage for later use. We refer the interested
reader for an analogous situation in a re-visit of Maxwell's demon
by \citep{key-7} to connect algorithmic randomness to physical entropy.
For the present purpose, it suffices to note that such a segregation
results effectively from an algorithmic feedback, requiring multiple
passes through the data before fusion. The output of this feedback
via multiple passes on stored data is also analogous to the concept
of a relay network with {}``side'' or meta information, described
in \citep{key-1-0}, where this side information is used by other
entities for selective processing of the data reaching them. In our
abstraction, a provision for such an algorithmic feedback should be
an essential part of the signal processing architecture, and should
be present in the path between the digitizer and the fusion operator.
This is only possible if the real-time system is equipped with adequate
memory and preprocessing for characterization of data before they
are sent to a correlator or beam forming system.\emph{ }By excluding
such a provision, digital receivers in existing large radio telescopes
are a potential source of irreversible biases in the recorded correlations,
and suffer from inefficiency for requirements like detection of short
term transients.

\section{\label{sec:RealTimeProcessing}Real-Time Processing Requirements}

Without getting into specific details of the real-time processing
required for a large multi-element radio telescope, we abstract them
into a combination of three broad categories:
\begin{itemize}
\item \emph{Embarrassingly Parallel processing}, e.g., spectral decomposition
of the incoming time series (say via FFT or polyphase filter banks)
and the recognition and management of path-induced distortions/interference
on timescales significantly smaller than the integration time chosen
for the correlations.
\item \emph{Pipelined processing}, e.g., multi-beam formation (say \textit{K}
beams) by phasing N elements requires $Klog_{2}N$ pipelined operations
for each spectral band.
\item \emph{Data Fusion }operations, in which the data originating from
different antenna elements are hierarchically fused (combined via
routing and application-dependent processing) along a chosen set of
dimensions which include time, frequency and spatial spread.
\end{itemize}
The most important fusion operation for an antenna array is the real-time
correlation of signals from every possible pair of antenna elements
in different frequency sub-bands. Apart from being an $O(N^{2})$
process (for an N element array) from a computational point of view,
this brings in the additional complication of routing large volumes
of distributed data to appropriate data processing elements to provide
a complete-graph connectivity between the sources of data and processing
elements. Significantly, cross-correlation between all possible pairs
of signals is also essential for using self-calibration techniques
to enable dynamic calibration of instrumental and atmospheric contributions
to the data corresponding to different elements, before they are subjected
to an irreversible fusing operation in a phased array. This makes
a spectral correlator an implicit requirement, even for a phased array,
for minimizing the irreversible loss of information resulting from
distortions induced in the the path or the local environment.

In our approach, we bifurcate the requirements of a real-time system
into \emph{Commodity} and \emph{Custom} segments. In the current state
of technology, the commodity segment can be fulfilled by subsystems
available in the market while the custom segment can generally be
realized on the basis of customized hardware and/or firmware layers
based on COTS technology. Such a bifurcation is explained below for
different functional categories of the NSPS:

\subsection{Computation\textbf{:}}
\begin{itemize}
\item \emph{COTS Segment}: Computationally complex and/or latency tolerant
processing, typically realized on a programmable platform ranging
from workstations to a high performance cluster.
\item \emph{Custom Segment}: Latency critical, logic intensive and repetitive
pattern of deterministic processing, well suited for a configurable
platform, typically FPGA-based. 
\end{itemize}
For efficient computation, we pay special attention to reducing coupling
between data in order to target explicit parallelism at all levels
of processing. Multiple parallel circuits are implemented in the FPGA-based
custom segment, while the current trend of multicore processors with
access to shared memory is exploited in high level software. Further,
the desired high signal bandwidth and large number of antenna elements
make the processing complex and compute intensive. This aspect, and
the advantage of quickly implementing exploratory algorithms, make
a commodity compute cluster an attractive choice for the central computing.
This is recognized by explicitly including the cluster in the COTS
segment mentioned above.

\subsection{Data Routing\textbf{:}}
\begin{itemize}
\item \emph{COTS Segment}: Commercial switches with all-to-all connectivity
are used for data routing to commodity processors and broadcasting
in the last mile, as well as for load balancing. The routing is controlled
by manipulating the destination addresses on data packets. Connection-less
protocols like User Datagram Protocol (UDP) are adequate for high-speed,
streaming applications where a small fraction of lost packets does
not affect performance adversely. Packet collisions are minimized
in full duplex, point-to-point connections between network partners,
and also because data flow is extremely asymmetric. Further, the criteria
for load partitioning discussed in Section \ref{sub:Load-Partitioning-and}
very often result in under-utilization of link speeds to match them
to sustainable processing bandwidths. 
\item \emph{Custom Segment}: Customized switches with static routes for
traffic shaping are relevant when only a subset of the network data
needs to flow to a subset of the nodes based on certain conditions.
They are generally implemented in configurable logic.
\end{itemize}
Since many FPGAs support Gigabit Ethernet MAC as a hard (or publicly
accessible soft) IP, this feature is useful while introducing a bridge
between the peer-to-peer network and the commodity network in the
last stage of the custom segment. In addition, some or all the major
subsystems may have management support from an embedded or explicit
on-board processor.

\subsection{Network Topology and Protocol\textbf{:}}
\begin{itemize}
\item \emph{COTS Segment}: A commodity network compatible with a typical
high performance compute cluster, which includes Gigabit Ethernet
as a \emph{de facto }standard for interfacing with external systems.
\item \emph{Custom Segment}: A peer-to-peer network which may include significant
on-the-fly application specific operations, suitable for implementing
on a standard FPGA platform. 
\end{itemize}
\begin{figure}
\includegraphics[scale=0.5]{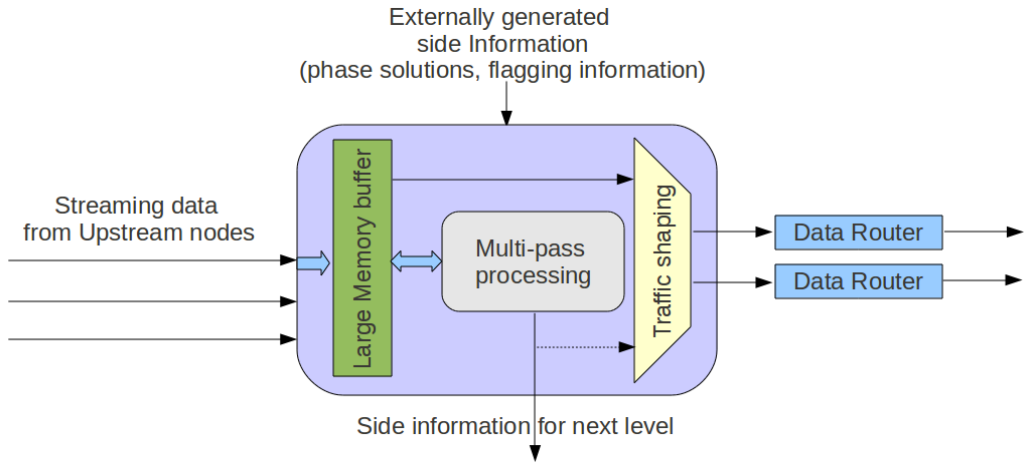}

\caption{\emph{\label{fig:A-Data-Pooler,}A Data Pooler node of the NSPS shown
carrying out data fusion and traffic shaping.}}

\end{figure}

An implementation of the actual processing on a dedicated set of identical
hardware circuits or parallel processors can take advantage of an
intelligent network capable of elementary on-the-fly operations to
achieve a balance on the dataflow and computing requirements. This
is depicted in Fig. \ref{fig:A-Data-Pooler,}, where a {}``Data Pooler''
node, as defined in Section \ref{sec:FusionTree}, is shown carrying
out real-time fusion of the streaming data by using the side information
made available by external sources. At the same time, the pooler is
seen generating side information out of the fused data set by way
of multiple processing passes on the stored data. The pooler can then
segregate the data, and route the segregated components to different
data sinks using the routing information available with the peer-to-peer
link nodes. We use {}``traffic shaping'' here in a more general
sense than in Internet traffic shaping (which delays lower priority
packets in favour of better network performance of higher priority
packets) to refer to both segregation of the incoming stream, as well
as the specific routing of segregated data to different sinks. Further,
the efficiency of hierarchical computation can be significantly improved
by accommodating some degree of pre-processing and/or partitioning
of data in each level of the custom segment to facilitate the next
level.

\subsection{Process Scheduling\textbf{:}}
\begin{itemize}
\item \emph{COTS Segment:} The overall task supervision, command and monitor,
user interface and the dynamic system monitoring are tasks whose complexity
is best left to the commodity segment to handle, where a variety of
tools ranging from MPI, compiler resources and advanced operating
systems like Linux or VxWorks are available.
\item \emph{Custom Segment}: Event driven scheduling with periodic or quasi-periodic
events generated conveniently in a low latency logic implementation
suitable for an FPGA. The interval between the events is stretched
to handle an application-specific transaction unit to the extent permissible
within the available resources.
\end{itemize}

\section{\label{sec:FusionTree}NSPS as a Fusion Tree }

In this Section, we present the NSPS architecture as a\emph{ Multi-sensor
Data Fusion Tree, }in which both conventional and {}``virtual''
sensors play a role. While entities like antenna elements, round-trip
phase/delay monitors, noise calibration etc can be treated as {}``conventional''
sensors, {}``virtual'' sensors result from processing blocks at
various levels. For instance, pre-processing can result in a flagging
mechanism to improve the reliability of fusion systems like correlators,in
which the original data are erased while compressing their information
into a statistical estimate to be passed to the next level.

\begin{figure*}[tp]
\center\includegraphics[scale=0.45]{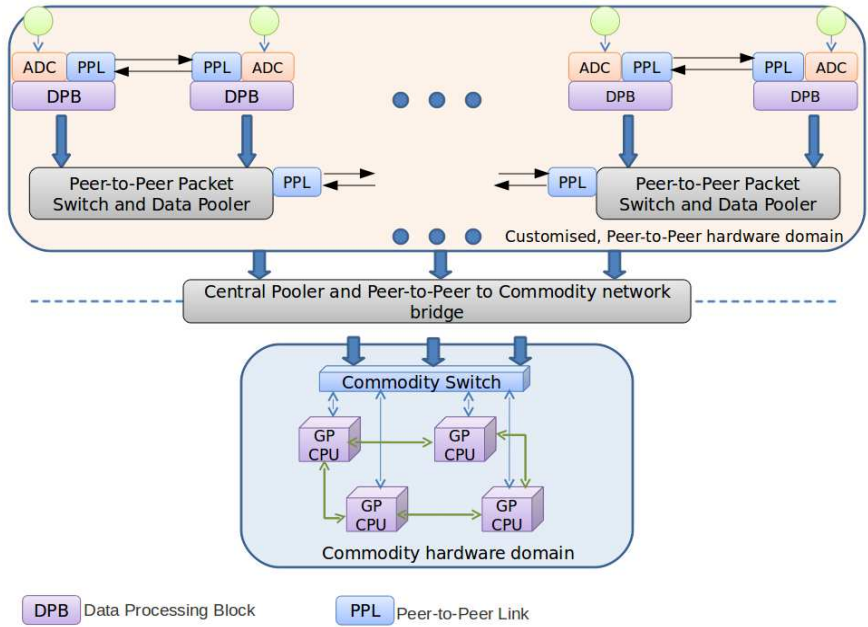}

\caption{\emph{\label{fig:A-conceptual-layout}A conceptual layout of a Networked
Signal Processing System architecture showing the principal participating
entities.}}

\end{figure*}

The signal processing system proposed in this paper is a set of spatially
separated nodes of varying communication and processing capabilities,
which are interconnected by a customized high speed tree-like packet
switched network interfaced to master commodity nodes. This is equivalent
to a \emph{Data Fusion Tree}, with the nodes of the tree performing
operations like traffic shaping, packet routing, or pre-processing
before data fusion. Accordingly, we have described the overall architecture
of NSPS in the form of a \emph{fusion tree}, schematically represented
in Fig. \ref{fig:A-conceptual-layout}. However, each level provides
a different mixture of functional capabilities. This has resulted
from our recognition of the following features of the functional requirements:
\begin{itemize}
\item Distributed computing across many nodes of different processing capabilities,
with local parameters guided by a processor capable of seeing a subset
of all data.
\item Data routing nodes with configurable routes for routing preprocessed
data subsets.
\item Nodes with large memory buffers to enable multiple passes on data,
also for enabling memory based data transposition.
\item A High speed network interconnecting all nodes, with ability of COTS
nodes to tap into the network.
\item Interface to a master commodity node through a standard interface
like Gigabit Ethernet (GigE).
\end{itemize}
Thus, from a functional point of view, we classify the nodes in NSPS
into the following three categories:
\begin{enumerate}
\item Data Poolers/Fusers: Nodes with sufficient on-board memory to allow
packetizing and multiple processing passes on incoming data. These
break the need for many-to-many connectivity in the correlation process
by transposing data via memory based switches. The transposition,
based on different parameters, ultimately serves a packet of data
suitable for processing by a single element of a distributed system
in an embarrassingly parallel manner.
\item Data Routers: These elements are endowed with high speed links to
either peers or more powerful processors to whom incoming packets
are routed based on statically configured routes. These form an integral
part of our architecture, helping in the load balancing by directing
appropriate subsets of preprocessed data to different processing elements.
We use commodity switches for routing data to multiple external sinks
by forming many-to-many connections between data sources and data
processors.
\item Data Processors: These elements have high compute density and can
be used for preprocessing, as well as for data rate reduction. We
classify processors into two groups as mentioned earlier:

\begin{enumerate}
\item Those catering to computationally complex and/or latency tolerant
processing, an example of which is the estimation of system calibration
parameters based on a long (few minutes) history of data, and its
dynamic updation. This processing is generally carried out by sending
a subset of the data to a central processor.
\item Latency critical, logic intensive and repetitive pattern of deterministic
processing. An example of this class is the real-time block-level
data encoding process requiring the estimation of block level statistics.
This can be carried out by multiple passes on small segments of data.
\end{enumerate}
\end{enumerate}
Thus, we visualize the NSPS as a \textit{restricted} distributed system,
depending primarily on stripped down lightweight networking protocols
and the static routes set up during system configuration. Data routers,
both customized and commodity, play an important role in reorganizing
data to be computationally palatable to processing nodes in this scheme.
A master node is in charge of command, configuration and control,
and is almost always a commodity node like a PC. It may be noted that
custom processing is spread across the NSPS tree by explicitly advocating
local intelligence in every node. As an illustration of the inherent
facilitation of distributed processing in the NSPS, some important
aspects of the interconnection mechanism are elaborated in the following
subsections. 

\subsection{Fusion tree interconnection model:}

Interconnects in the Data Fusion Tree consist of the following four
essential graphs, which can either be logical or explicitly physical
manifestations.
\begin{enumerate}
\item The \emph{Data} network is a simplex, high bandwidth net connecting
the leaf nodes of the Fusion Tree to a central processor, possibly
passing through several collation levels of the signal processing
tree.
\item The \emph{Control and Monitor} network is a full duplex, low bandwidth
network and interconnects all nodes hierarchically through management
processors to a central monitoring station.
\item The \emph{Calib} network is a full duplex, low bandwidth network.
This allows calibration information to reach the data fusing nodes
before the sequence of irreversible fusion operations take place.
\item The \emph{Clock} network is a full duplex network, providing the distribution
of clocking signals to the various nodes, as well as allowing a round
trip clock phase measurement.
\end{enumerate}
In actual implementations, it may be simpler to realize these in terms
of a set of simplex networks, among which clock, control, monitor
and calib are directed towards the leaf nodes while the data and status
(including response to monitor queries) belong to simplex networks
which flow from different levels of the fusion tree into the master
node. 

Our network implementation can be bifurcated into the following sets: 
\begin{enumerate}
\item Customized high speed serial peer-to-peer links terminating into peer-to-peer
switches which implement a subset of the complete graph connectivity.
\item Commodity high speed serial links terminating into commodity networking
equipment, with ability to interface to standard processing nodes. 
\end{enumerate}
In typical implementations, we expect custom links to be on a Passive
Optical Network (PON) based communication stack. Gigabit Ethernet
is the preferred choice for the backbone of commodity segment to connect
to the custom network.

We exploit the high speed serializing ability of modern FPGAs to dispatch
data on high bandwidth copper links and use PON components to meet
the spatial spread required to reach remote nodes over fibre links.
As long as the bandwidth requirements are met, no specific preference
is implied for a choice among different networking technologies. Thus,
some implementations may utilize the embedded multi-gigabit serializers
in FPGAs for peer-to-peer links while others may refer the cross dispersion
of data to a compute cluster's high bandwidth infiniband network,
or use commodity Gigabit Ethernet. This leads to the need for a flexible
bridging mechanism which can be exploited by implementations. For
instance, data can be conveniently transmitted over a peer-to-peer
or a commodity link due to the maintained commonality of their interfaces.

\subsection{NSPS tree network characteristics:}

The high speed network internal to the NSPS tree has features which
are restricted and stripped down versions of those found in commodity
networks. This is an optimization due to the highly controlled network
which exists within the telescope receiver environment. Our network
differs from regular networks in several aspects:
\begin{itemize}
\item A controller node is assigned for every sub-tree at a given level.
This entity forwards command and status information between controller
(level-0) and upstream levels. Thus, it is not a typical peer-to-peer
network which does not have such a hierarchical control structure.
Broadcast and multicast domains available in commodity networks are
used to implement control and monitor mechanisms, while explicit {}``pull''
mechanisms are implemented on the custom network. Here, the {}``pull''
refers to the explicit request for data made by a downstream node
to an upstream node. The advantage of a {}``pull'' mechanism is
that data is made available to a downstream node only when it is ready
to handle the data, as inferred from the downstream node's request.
Also, if a downstream node is busy, then the upstream node loses data
in integral packets, thus maintaining timing information.
\item The network configuration and routes are fixed statically in an application
dependent manner at configure time. There is no node discovery, and
data routing does not have an explicit mechanism for handling node
failure. Since all data flows towards a logical sink, there is also
no destination address, although source addresses can be preserved.
This simplifies network management to a large extent at the cost of
non-redundancy of NSPS entities.
\item Communication protocols: All elements in our network, including master
nodes, generate similar kinds of packets which contain an 8 byte header
with fixed fields. These are typically indicative of the nature of
accompanying data, as well as its timestamp, source and other meta
information. System state can also be propagated through these packets,
or by forming special status packets. The restricted meta information
processing makes it simpler to realize packet formation in hardware
with simple state machines.
\item High speed serial interconnects: All entities in the internal network
communicate via high speed serial interconnects with a clock recoverable
from the encoded data. This approach allows us to transmit data long-haul
over fibre, or short-haul over copper without any changes. In particular,
we discourage bus-based interconnection between physically separated
nodes.
\end{itemize}

\subsection{Interface to external network:}

Once preprocessed data is ready within the NSPS, it needs to be transferred
onto a commodity network for reaching commodity nodes for post processing
or archiving. Local intelligence in the peer network can be used to
partition the data such that the interfaces to commodity nodes use
link speeds commensurate with their processing ability. For simplicity,
we have used Gigabit Ethernet as a typical standard external interface.
This is a popular high speed serial interconnect with a vast amount
of infrastructure available in the commodity market. It also allows
transmission over copper (UTP) to interface directly with commodity
servers, or fibre (via conversion to 1000BaseX) for long-haul transmission.
Commodity servers of moderate ability can then be used as data sinks
with minimal customization. This is also motivated by the fact that
many modern FPGAs have embedded high speed serial interconnects on
chip, with complete Gigabit Ethernet support in the form of on-chip
Gigabit MACs or as publicly available libraries.

For the last mile connectivity, UDP can be used since it is a simple
connectionless protocol with minimal overheads on top of IP. It is
also possible to fill the relevant UDP fields during system configuration,
and hold them static for the duration of an observation. Each of the
internal network types carrying data (\emph{Data, Calib }and\emph{
Control}) can then be easily made available over a different UDP port
as part of the design. This allows an application program to associate
independent threads to service these streams.

\subsection{Control and Monitoring network:}

The distributed nature of our architecture requires status monitoring
of all nodes and links, which can be handled by the individual sub-tree
roots and communicated to the master controller. This is implemented
by a status {}``pull'' scheme by which controlling entities periodically
query the status of all nodes in the NSPS tree rooted with them by
way of a special AYA ({}``AreYouAlive'') packet. The nodes respond
with an IAA ({}``IAmAlive'') packet containing selected status information.
Similarly, control packets contain command and configuration data.
Each command packet typically results in a status reply from the targeted
entity, which confirms the receipt of the command, and regenerates
a control packet for the entities controlled by it. The master node
can use this in an appropriately scheduled housekeeping operation
to discover failure of nodes.

\subsection{Calibration network:}

This network is meant to carry data from the NSPS tree which is relevant
to forming calibration solutions for the array. The calibration mechanism
is to be applied differently for the two main modes of observation
with the NSPS:
\begin{itemize}
\item In the interferometric mode, the correlator can work independently
of the actual gains and phases of the sensor elements, since the observations
include calibration scans at reasonable time intervals. Off-line processing
can infer intermediate variations by supplementing interpolation between
calibration scans with dynamic calibration schemes like self-calibration
based on the partial, low bandwidth dataset available over the calibration
network.
\item On the other hand, real-time beam formation includes an irreversible
fusing operation which requires dynamic calibration to be part of
data fusion. Fortunately, it is often possible to use a relatively
small subset of the data (non-contiguous timeslices or a chosen frequency
sub-band) for this purpose to enable short term predictions of gain
variations. These can be fed to the fusing nodes well in time before
irreversible fusing operations are performed. Since the complexity
of the actual algorithm used for calibration makes it better suited
for a general purpose computer, the \emph{calibration network} can
be used to route the relevant subset of data to a commodity switch
and deliver the calibration parameters to the appropriate NSPS tree
level.
\end{itemize}

\subsection{The clock network:}

In the spatially distributed, direct RF sampling NSPS architecture,
clocks passed to samplers have very stringent signal quality constraints
in terms of net jitter and stability. The alignment of the multiple
data sources before fusion requires high relative stability of the
sampling clocks with random jitters much smaller than the reciprocal
of the highest frequency in the sampled signal. Clock distribution
should also include a mechanism for ensuring the traceability of timekeeping
at all digitizing blocks to a centrally maintained time standard to
a very high accuracy. The implementation can benefit from commercial
clock distributors which have embedded phase-lock loop clock synthesizer
with on-chip Voltage Controlled Oscillator (VCO) and a per port delay
tuning for the distributed clocks.

\subsection{Data Network:}

All data flowing in the NSPS is packetized with a custom, low overhead
header. All subsystems accept and generate data in a packetized fashion.
This reflects the inherent asynchronous nature of our system. Packets
traversing our platform are atomic and capable of independent existence.
The packetizing of data means that data loss due to network congestion
or buffer over-runs is never arbitrary, but always in units of packets.
At any instant, our network can have different kinds of packets traversing
it, corresponding to different stages in the processing. The basic
unit of packet size is maintained as 8 bytes, which is a natural unit
or sub-unit for different memory and processing hierarchies. Adequate
padding is used if necessary to maintain this condition. The header
is mandated to have a few fixed fields which are common in size and
layout across packet types, allowing processing entities to examine
packets which can be processed by them, while discarding the others.
In a broadcast network, this approach can waste bandwidth when packets
are discarded, but the wastage can be minimised by setting up static
routes between partner nodes. This is possible in both the custom
and commodity peer-to-peer link nodes. The Command network, on the
other hand, is a broadcast network, with nodes passing on commands
not addressed to them to all other nodes downstream of themselves.
Data sources can include packet specific extensions to the packet
headers generated by them. The following fields are suggested as a
mandatory part of the packet header:
\begin{itemize}
\item Source identifier: At every level of the tree, nodes are endowed with
a unique identifier which supplants the existing upstream source id,
if any processing is carried out on the packet.
\item Datatype: This field allows processing entities to recognize which
packets are palatable to them and to reject others. 
\item Data pixel descriptor: This field lays out the size and description
of the smallest unit of data transfer to be one of an allowed set,
which is implementation dependent.
\item Streams: This field records the number of independent signal sources
present in each packet.
\item Packet size: The size of a packet is expressed in units of words as
specified by the datatype field. 
\item Timestamp: This field is populated as early as possible in the data
generation path and maintained across data processing. This field
is generally populated by a timestamp counter running on either a
reference clock or on the sampling clock itself and is traceable to
the centrally maintained time standard.
\end{itemize}
This specification is efficient for real-time streaming data description
with minimum overhead. For archival of processed data, a standard
format which allows multiple binary streams to maintain their identity,
like the FITS or VLBI Data Interchange Format (as proposed by the
VDIF Task Force (2009)) can be used.

\subsection{\label{sub:Load-Partitioning-and}Load Partitioning and Scheduling:}

No subsystem in our scheme is source synchronous, be it at the hardware
or the software level. All subsystems have enough memory for a store-and-forward
of several packets. This allows the processing to happen at the packet
level, on a faster clock than the sampling clock of front end ADCs.
It also eases the timing requirements of designs implemented in FPGAs
and makes them more tolerant of clocking errors. Sequencers play an
important role in our architecture, generating events on which the
processing progresses. The sequencers generate necessary globally
aligned events to which any action taken on the basis of commands
from commodity network will get aligned. A simple example is an implementation
where all processes in the peer network operate at a block level,
with a periodic event signifying the need for scheduling a new process
as a result of the arrival of a new block of data. 

It is important to match the communication bandwidth to processing
abilities at every level in the signal processing tree. More specifically,
event markers generated by the sequencers should facilitate a partitioning
of processing in each level into abstract transactions, where each
transaction deals with the entire data collected over a convenient
timeslice \textbf{and} the relevant data are locally available \emph{on
demand}. In particular, it is desirable that processing at the central
commodity segment is facilitated at a cadence suited for general purpose
operating system scheduling to achieve latency tolerance. For instance,
to be commensurate with a housekeeping tick of 10 milliseconds in
typical Linux configurations, the transaction timeslices should be
several times longer.

\subsection{Choice of fusing dimension:}

Among the available axes in processing space along which the data
can be partitioned and distributed to parallel processors, the time
axis is often the most convenient for slicing, as individual timeslices
can be considered independent. For a large network, we realize this
from a hierarchical set of Data Poolers\emph{ }populating different
levels of the processing tree, which can collate data from different
sources and partition them along the time dimension at each level.

\section{\label{sec:An-Implementation-Example:}An Implementation Example:
Reconfiguration Plan for the Ooty Radio Telescope }

In this section, we provide an example of implementation of NSPS by
giving an outline of the system being planned for modernizing the
ORT \citep{key-5}. The ORT is a 506m X 30m equatorially mounted cylindrical
telescope with an equispaced linear array of 1056 dipoles along the
focal line. Each dipole has a tuned low noise amplifier \citep{key-4}
with about 40 MHz bandwidth centered at 327MHz, although the existing
analog phasing network restricts the bandwidth to about 10 MHz. A
collaborative program for upgrading the ORT has been undertaken jointly
by the Raman Research Institute (RRI) and the National Centre for
Radio Astrophysics (NCRA), which operates the ORT. In this program,
the feed array is logically divided into 264 identical segments, where
each segment represents an independent antenna element of size 1.93m
x 30m. The aim of this upgrade is to reconfigure the ORT into a programmable
264-element array. When completed, the reconfigured ORT will have
an instantaneous field of view of $\sim27^{o}$, bandwidth of $\sim35MHz$
and will be equipped with an NSPS-based digital receiver. Currently,
prototypes have been tested for a large fraction of the custom segment
of the NSPS and the analog signal conditioning subsystems. The final
production and integration is expected to be completed in 2011. The
digitizers are organized in 22 digitization blocks located below the
reflector at a spacing of about 23m, where each digitization block
includes a 12-channel digitizer capable of operating at 100 Ms/s.
All the 22 digitization blocks are connected in a star topology with
a central system using a peer-to-peer network on optical fibres with
multiple links operating at speeds of 2.5 Gigabits/sec from each block.

\begin{figure}
\includegraphics[scale=0.43]{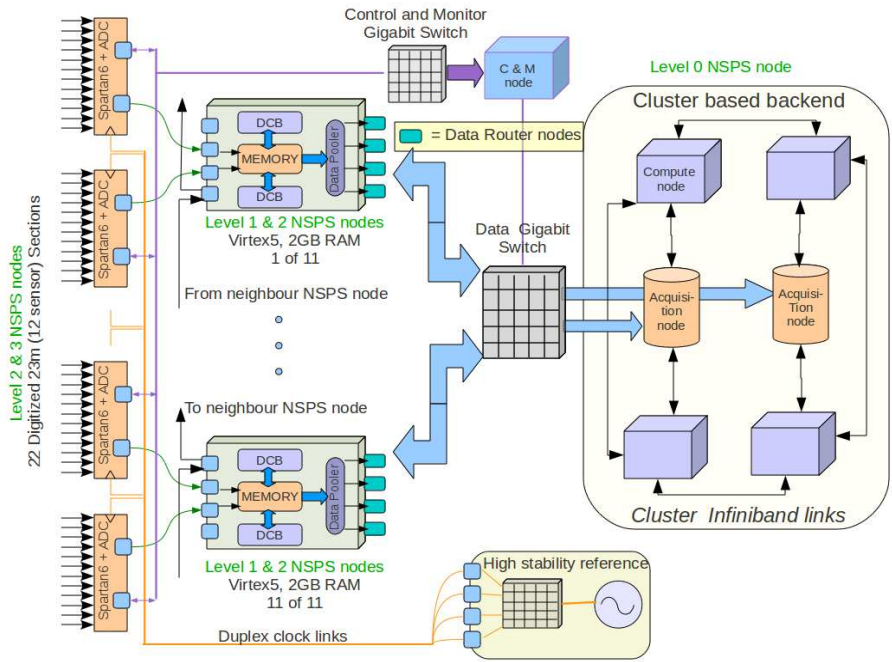}

\caption{\emph{\label{fig:Proposed-NSPS-implementation}Proposed NSPS implementation
for configuring ORT as a 264 element programmable telescope.}}

\end{figure}

\subsection{The ORT NSPS layout:}

The proposed system consists of four hierarchical levels as illustrated
in Fig. \ref{fig:Proposed-NSPS-implementation},\textcolor{black}{{}
where level-0 is the root node and realized by a high performance
cluster with Infiniband for inter-node communication and GigE for
communication with the NSPS.} The high speed peer network uses the
light weight Aurora protocol simplex links for data uplink, with last
mile via GigE. Currently, bandwidths of upto 100MBps per GigE link
into level-0 memory have been sustained.
\begin{itemize}
\item \emph{Level 3:} This level is composed of the distributed digitization
infrastructure and is installed at the antenna base. The prime components
of this level are the digitizers, the sampling clock derivation and
conditioning circuitry, the first level data organizer and peer-to-peer
link handler. All these components required for handling 12 sensor
elements from a 23m section of the ORT are implemented as a single
board.

\begin{itemize}
\item The ADCs (dual channel AD9600) are capable of a 100Ms/s sampling for
signals with frequencies upto $\sim600MHz$. The $\sim60dB$ dynamic
range at 10 bit resolution allows us to implement an AGC in software.
More importantly, the implementation can benefit from the on-chip
sampling clock conditioner, divider and duty cycle stabilizer. For
instance it is possible to provide a sine-wave with frequency 2-8
times the sampling clock, and use the on-chip features to convert
to square-wave, enable duty-cycle stabilizer and divide by suitable
integer to get the sampling clock, thus reducing the overall sampling
clock jitter and hence the phase noise in the sampled data. This feature
is useful in direct (harmonic) sampling of the incoming $327MHz$
RF since the Nyquist sampling interval in a band-limited RF is decided
only by the bandwidth while jitter tolerance depends on the highest
frequency content.
\item This level has an embedded clock synthesizer and distributor for the
on-board 12-channel ADC based on a reference distributed on fibre
by the central high stability clock distributor. It is received using
a digital fibre optic receiver. The embedded clock distributor is
based on a clock buffer and distributor (LMK03020) which has an on-chip
VCO and per-port delay tuning.
\item The first Data processing block is implemented in a Xilinx Spartan6
(LX45T) FPGA to perform a conversion of the ADC 10 bit resolution
data to $\sim4$bit via configurable, table-driven logic, where the
look-up-table (LUT) is dynamically updated to accommodate innovative
schemes of compression and segregation. For instance, let us assume
that a choice among a pre-determined set of LUTs is best suited for
coding/compressing the data in a set of physical packets associated
with a transaction unit. Here, each table implements a different encoding
of the input word to an output with lesser number of bits per word.
Further, we assume that every word of each physical packet is encoded
by a choice between two LUTs out of the set, to represent normal and
segregated(flagged) data. For decompression by downstream nodes, the
tags of these two LUTs can be accommodated in the packet header, while
a one-bit selection between the two can be associated with each data
word - thus achieving the dual purpose of flagging and scaling at
the word level. Such a scheme can accommodate a wide range of scale
factors and hence a large dynamic range within a logical packet. Suitable
thresholds for packet-level choice of LUTs can be generated on the
basis of integrated power over a reasonable time stretch as part of
the pre-processing.
\item The Data Router node buffers data from all 12 sensors in internal
memory and reorganizes them to form packets containing identical time-stretches
from all antenna elements. The peer-to-peer link out of each 23m section
which connects level-3 to level-2 is implemented using the 4 available
RocketIO multi-gigabit onboard serializers on the Spartan6. Our choice
of SFP for the current implementation can sustain link speeds of upto
2.6Gbps on single mode fibre, while we use the light-weight Aurora
protocol at a wire speed of 2.5Gbps for communication between level
3 and level 2.
\end{itemize}
\item \emph{Level 2:} This level is implemented using an FPGA (Virtex5 LX50T)
board whose on-board resources include 2GB of memory and 8 multi-gigabit
transceivers and expansion connectors. This board can sustain the
following level-2 functionalities:

\begin{itemize}
\item FFT block: Here, the data processor block first decodes data from
a pair of sensors, packs them into the real and imaginary parts of
a 32-bit complex integer word, and implements a pipeline stage (e.g.,
radix-64) of a split radix FFT for all pairs of incoming channels.
The processing resources are enough to handle upto 24 channels (2
level-3 entities) at the maximum sampling rate. 
\item The Data Pooler block operates using the large local DDR2 RAM and
the interconnection with other level-2 cards to pool subsets of both
local and remote level-2 data into transaction oriented packets by
partitioning data along the time axis. Here, each transaction refers
to the processing of a specific timeslice for the entire array.
\item The Data router block collects data from the local RAM in units appropriate
for transfer to each outgoing port, packetizes them and sends out
selected time slices to level-1 entities. Provision for computation
offloading is provided in the form of spare peer-to-peer links which
can add on more level-2 cards.
\end{itemize}
\item \emph{Level 1: }At this level, a memory-based, NSPS to commodity network
bridge is implemented. Large bursts of continuous time slices are
first buffered in RAM, and then sent out over GigE as properly timestamped
packets. The level implements load partitioning as configured by the
root level by manipulating ethernet destination addresses of streams
going into the data GigE switch. There is possibility of implementing
a data processor block for the remaining 16 point FFT operation pending
from the split radix FFT.
\item \emph{Level 0: }At the root of the NSPS tree, a medium level cluster
is proposed to handle both the communication and processing requirements
for forming correlations between all sensors. It is important to note
that the cluster inter-node traffic is significantly reduced due to
the data routing and transposition carried out using the Data Pooling
nodes at the various levels. The formation of actual correlations
and the calibration parameter estimation is carried out by this level.
The above mentioned partitioning of the load into the 3 levels can
be used to bring a subset of data from all 264 elements into one node
via a quad-GigE card.
\end{itemize}

\subsection{System Control:}

The Control network is a simplex, one-way channel from a master with
unique id (in the commodity segment) to the peer network through the
bridge node. Thus, while data can be routed to arbitrary nodes in
the cluster depending on the UDP destination addresses set during
configuration, commands are accepted by the bridge node only through
a privileged link from the master. This simplifies assigning privileges
to operations related to starting, stopping or resetting the acquisition
state machines, configuring the network routes on customized hardware
switches, changing ethernet destination addresses, or changing the
contents of the LUTs used in earlier nodes like the digitizers.

\section{\label{sec:discussion}Discussion}

While asynchronous, packetized processing over standard networks is
a relatively new concept in radio telescopes, it is being embraced
enthusiastically due to the many benefits it offers to the system
designer. Even among this class of telescope data processors, contemporary
architectures usually have a direct link from the samplers to the
central processor. Some operations like a digital filter bank or FFT
are carried out remotely, while others like cross-correlation is done
centrally. The memory-rich architectures of modern FPGAs help in distributing
computing to remote nodes and enables buffering to allow multiple
passes on the streaming data. As the data volume grows, e.g., in the
central pooling stations, the processing can be supported by large
off-chip memory using commercial memory modules, routinely supported
by modern FPGAs. This provides substantial enhancement to buffering
for transaction level operations and data partitioning. A peak data
rate of 100 Ms/s x 4bits for 264 elements for the ORT would correspond
to about 13.2GB/s for which the level-1 buffering of 22 GB in 11 Virtex-5
cards shown in Fig. \ref{fig:Proposed-NSPS-implementation} is comfortable
to sustain transactions of upto a fraction of a second duration. The
use of standard software stacks also allows us to leverage the various
high performance modes being worked upon by system optimizers, e.g.,
the zero copy mechanism on Linux. 

Another challenging problem with large arrays is the so-called \emph{corner
turning }problem\emph{, }which refers to the transposition of the
input signal matrix needed to achieve the all-to-all communication
necessary for correlation. Earlier approaches have looked at either
commercial switches or entirely customized switches for routing data
\citep{key-1}. We break this problem down into levels, and apply
a hybrid of commercial as well as custom routing. The Data Pooler
element is utilized to implement a memory based switch, while the
COTS (GigE) network controller manages another level of redirection
by manipulating UDP destination addresses.

\section{Conclusions}

We have presented a packetized, heterogeneous and distributed signal
processing architecture for radio interferometric signal processing
which elevates the network to a core system component. The architecture
addresses some of the core issues pertaining to interferometric signal
processing. We visualize this problem as that of an appropriate workload
creation and scheduled dispatch to matched processors over a data
flow tree. Here, the leaf nodes are sources of data, with data processors
handling a managed slice of the processing at the intermediate nodes
of this tree. We emphasize the use of COTS components, both hardware
and software, for rapid deployment, ease of maintenance, and lowering
the cost of the architecture implementation. 

The goal of realizing a programmable telescope with NSPS is facilitated
by defining rigid interfaces between both hardware and software components.
This can allow exchange of a variety of data with varying communication
and computing requirements between levels in the network. Most of
the individual nodes in the NSPS can change the nature of their processing
within the limits specified by their designed personality and available
resources at the node. This allows offloading of computing requirements
in a hierarchical manner up the NSPS tree, trading off implementation
time with hardware capability of an application mode. Due to the rigidity
of interfacing protocols as well as the standardized networks making
up the system, we can comfortably add nodes which can tap into the
NSPS in order to carry out a different processing chain. Data duplication,
if required, can be carried out by COTS components (e.g., by switches
operating in broadcast mode) thus reducing development load. 

We have presented an outline of the NSPS implementation being planned
for configuring the ORT as a programmable 264 element telescope. Our
architecture is optimally tuned to service the needs of medium sized
arrays. We advocate full software processing for smaller arrays, with
an increasing factor of hardware offload as the array size grows.
This approach has being taken by us in building a 44 element demonstrator
as a precursor to the receiver for the full 264 element ORT array.
This receiver exploits all NSPS aspects we have dwelt on, and is in
an advanced stage of completion.
\begin{acknowledgements}
The 44-element demonstrator for the ORT is being built as a collaborative
effort between Radio Astronomy Laboratory at Raman Research Institute
and the observatory staff at the Ooty Radio Telescope. Many of the
ideas presented here evolved during the trials of demonstrator subsystems
for which we specially thank colleagues both at RRI and ORT. CRS would
like to thank Madan Rao and Dwarakanath whose comments have helped
improving the manuscript. We thank the anonymous referee whose comments
have been very helpful in improving the clarity of presentation in
the manuscript\end{acknowledgements}

\end{document}